\documentclass{article}
\usepackage{graphicx,graphics}
\title{\bf Real inertons against hypothetical gravitons.
           Experimental proof of the existence of inertons\footnote{
           {\it Indian Journal of Theoretical Physics}
           {\bf 48}, no. 1, pp. 1--23 (2000)}}

\author{{\bf Volodymyr Krasnoholovets and Valery Byckov} \\
{} \\
  Institute of Physics, National Academy of Sciences,  \\
  Prospect Nauky 46, UA-03028 Ky\"{\i}v, Ukraine}

\date{October 1998 -- November 1999}

\begin{document}
\maketitle

\begin{abstract}
Previously, one of the authors  has suggested [{\it Phys. Essays}
{\bf 6}, 554 (1993);  {\bf 10}, 407 (1997); xxx.lanl.gov
quant-ph/9906091] a mechanism of the particle motion within the
framework of a vacuum regarded as an original cellular medium,
i.e. quantum aether. The existence of special elementary
excitations of the aether medium  -- inertons -- around the
particle has been the main peculiarity of that mechanism. The
present paper  treats the impact of inertons on the collective
behaviour of  atoms  in  a solid.  It is shown that inertons
should contribute to the effective potential of interaction  of
atoms  in  the  crystal lattice. The possibility of separating
this inerton contribution from the value of the atom vibration
amplitude is analysed. The experiment which assumes the presence
of the hypothetical  inerton field is performed. The expected
changes  in  the structure  of the test specimens caused by this
field are in fact convincingly fixed in micrographs.   \\

\hangindent=2.9cm {\bf Key words}: space, matter waves, inerton
field, condensed matter, lattice vibrations, morphological
structure

\vspace{2mm} \hangindent=2.2cm {\bf PACS:} \ \ \ 03.75.-b Matter
waves -- 43.25.+y Nonlinear acoustic  --  14.80.-j Other particles
(including hypothetical)

\end{abstract}

\newpage

\section{Introduction}
\vspace{2mm} \hspace*{\parindent}
    Gravitational waves of the general theory of relativity
are presumed to be real carriers of the gravitational interaction.
The waves came out, based on the belief that the matter was highly
rarefied, Bergmann [1]: space between distant massive points was
perceived to be empty and hence a moving massive point did not
interact with a vacuum. It is assumed that gravitational waves are
quantized and that quanta of the gravitational field -- gravitons
-- are {\rm {\underline {massless}}} particles (see, e.g. Weinberg
[2]). However, it is important to keep in remember that Einstein
[3] derived the solutions for gravitational waves in 1918 and
those solutions were resting on the pure classical consideration.
Quantum mechanics was constructed later in the mid-1920s. It
introduced such a notion as the matter waves which are described
by the $\psi$-wave function. And just these waves as we know are
of primary importance in determining the behaviour of matter at
the scale comparable with the de Broglie wavelength $\lambda$ of
the quantum system studied. On the atom size at which the means of
quantum mechanics operates the $\psi$-wave function is considered
as the probabilistic characteristic of the quantum system. The
Schr\"odinger nonrelativistic formalism and the Dirac relativistic
one were developed just for the description of steady-states of
the electron in an atom in which the de Broglie wavelength
$\lambda$ defines the length of stationary orbit of the
corresponding electron. In the range lesser $\lambda$ the
behaviour and construction of the matter waves have not
practically been investigated, de Broglie [4].

     Turning back to the gravitation we should note that it has never
taken into account the presence of the matter waves. In this
connection one can raise the question whether the hypothesis of
the theory of gravitation about the existence of a macroscopic
quantization of space is correct. Probably it is quite reasonable
to assume that the theory is based on an essential methodological
mistake. Evidently, the main error was made when gravitational
waves deduced in the framework of phenomenological theory were
quantized in a scale much exceeded the de Broglie wavelength.  In
other words, the large-scale quantization was thrown on space
without any investigation whether such a requantization could
exist on the background of reliably established quantization in
the microworld. Therefore, in this case one faces a conflict
between the general theory of relativity and quantum mechanics.
Other significance aspects of the conflict have been discussed in
literature as well (see, e.g. Stapp [5]). Moreover, the background
of general relativity itself can be reconsidered from new
standpoints, Kar [6] (see also Logunov and Metvirishvili [7]).

    Modern theories of gravitation try to reach quantum foundations
starting from the macroscopic background. However, we can attempt
to construct the gravitation phenomenon beginning with quantum
mechanics. In fact, one can suppose that the matter waves which
are associated with structural blocks (electrons, nuclei, atoms,
{\it etc.}) of an object overlap and form the whole matter field
surrounding the object. It is apparent that a solution of this
difficult problem will be possible only in the case when we clear
up the substructure of matter waves.

    Currently several new views on the nature of a vacuum
substance have been introduced in quantum physics. Among them one
can point out mathematical approaches by Hoyle [8] on the origin
of matter, by Kohler [9] on point particles as defects in solid
continuum, and Bounias and Bonaly [10,11] who consider the matter
as deformations of space and in particular, could find
relativistic principles (the Lorentz constriction) and investigate
the gravity and quantum properties in the framework of the
topology and the set theory. There are also works which consider
the vacuum as a special kind of crystal, Fomin [12], as a "liquid
crystal" model, Aspden [13], as a zero point energy field, Haisch
{\it et al.} [14], as an electromagnetic energy density, Vegt
[15], as an inelastic Planck aether, Winterberg [16] and as a
degenerate electron-positron plasma that dominates the aether,
Rothwarf [17]. Winterberg's Planck aether consists of positive and
negative Planck masses, which interact locally through
contact-type delta function potentials. It is interesting that
these potentials are neither electromagnetic nor gravitational.
That does Winterberg provide an in-depth analysis has shown that
in microcosmos a new type of physical interaction could quite
exist.

     Previously one of the authors studied [18-20] the motion of a
canonical particle in a vacuum regarded as a cellular elastic
space. In this approach space is superdensely packed with
structural units, or cells -- superparticles -- which are found in
a degenerate state and whose size is of the order of $10^{-28}$ cm
(all kinds of interactions come together at this scale, as
required by the grand unification of interactions).  It was
conceived that a moving particle was interacting with
superparticles of the space net and as a consequence elementary
excitations called "inertons" were knocked out of the particle.
These excitations are virtual, since each emitted inerton is again
absorbed by the particle.  So a moving particle is constantly
surrounded by a cloud of oscillating inertons and this oscillating
nature of the motion is also applied to the particle. It has been
found that this kind of motion is characterised by the two basic
quantum mechanical relations:  $E=h \nu$ and $p_0=h/\lambda$ where
$p_0$ is the initial momentum of the particle and $\nu$ and
$\lambda$ are respectively the frequency and amplitude of spatial
oscillations of the particle (to put in differently, these are
peculiar submicroscopic overdeterminations of the de Broglie
wavelength $\lambda$ and the frequency $\nu$.)  As is well known,
de Broglie$^{21}$, the availability of these two relations enables
the wave $\psi$-function to be introduced which in its turn
results in the Schr\"odinger wave equation. The relationship
between the parameters of the particle and the cloud of inertons
has the form [18]
\begin{equation} v_0 /\lambda =
c / \Lambda.  \label{1}
\end{equation}
where $v_0$  is the initial velocity of the particle, $c$  is the
initial velocity of inertons (speed of light) and $\Lambda/\pi$ is
the enveloping amplitude of the inerton cloud which oscillates in
the neighbourhood of the particle.  Consequently, the particle's
inerton cloud is extended on a distance of $\lambda$ along the
particle path and restricted by the size $2\Lambda /\pi$ in the
transversal directions. Such motion of the particle studied from
the submicroscopic deterministic view-point is easily result in
the Schr\"odinger and Dirac formalism at the atom range [18-20].

    It is known that a behaviour of each element
of a solid such as the electron or a more composite system
(nucleus, atom) is characterised by its own de Broglie wavelength
$\lambda$ and own wave $\psi$-function. Based on our concept of
space we can consider any solid as being built into the space net.
In this case one can suggest that it is this inner substance, that
exerts control over electrons and atoms in solids, in the same way
as space governs a free moving elementary particle [18,19]. The
quantity $\lambda$ of the composite system clearly characterises
the coherent motion of all strong-coupling elementary particles of
which the system is composed. Therefore, inerton clouds of
separate particles merge into a common inerton cloud of the moving
system and then relation (1) can be applied to this system as
well. The application of the wave $\psi$-function to the
description of the whole composite system should mean that any
moving atom/nucleus is shrouded in an inerton cloud much as a free
moving elementary particle.

    It has been recently demonstrated by the author [22] that clouds
of inertons enclosing electrons reveal themselves in a great
number of experiments. Thus based on our results [18-20,22] we can
say that the wave $\psi$-function of any quantum system, from a
particle to a solid, is "filled" by a huge number of very light
inertons. This means that one may speculate that inertons could be
emanated from the $\psi$-function, that is, from the system's
inerton cloud and, because of this, they could be fixed by the
instrument. The nature of inertons is not directly associated with
the electromagnetism or gravitation. The inertons are more likely
to belong to the nature of matter, that is to the space net as it
follows from our concept (and the same possibility lies  in the
Winterberg's aether model [16] as well). Moreover one can infer
that just these elementary excitations of space may successfully
substitute for hypothetical gravitons of the general theory of
relativity (recall that the latter are derived from a pure
classical consideration, see e.g. Refs. [1,2]).

     The goal of the present work is to study the impact of inertons
on the collective behaviour of atoms in a solid. The problem is
regarded both theoretically and experimentally. In the first part
of the work we show how inerton clouds of separate atoms, which
are overlapped in a solid, form a system of entirely cooperated
inertons that is interpreted as a field of the matter waves of the
body studied. A contribution of inertons to the atoms vibration is
considered in the framework of a modified standard model of the
harmonic interaction of atoms in the crystal lattice. In the
second part of the work an experiment which concerns the
possibility of an inerton field to act upon the test specimen is
carried out. The microstructure of the reference and test
specimens is investigated using the electron microscope. The
corresponding micrographs appear in the affixed figures.

\vspace{4mm}

\section {Inerton contribution to crystal atoms \break vibration}
\vspace{2mm} \hspace*{\parindent}
     Atoms in solids vibrate  with  respect to the equilibrium positions.
Let us evaluate the velocity of the  atoms passing through the equilibrium
position, based on the equality of kinetic and thermal energy
\begin{equation}
Mv_0^2  / 2 \approx  k_{\rm B}T . \label{2}
\end{equation}
 At the room  temperature, assuming,  e.g., for the atomic mass $M= 30
 M_p$, where    $ M_p = 1.67\times 10^{-27}$ kg   is the mass  of
proton at rest, we find from (2) the  value  of the velocity:
$v_0\simeq 4 \times 10^2$   m/s.  The respective  de Broglie wavelength is
$\lambda  = h / M v_0 \simeq 3.3 \times 10^{-11}$ m.
According to our concept any motion of atoms in space should be
accompanied by the motion of inertons. Hence, substituting  the atoms'
values $v_0$  and  $\lambda$ into expression (1) we obtain for the amplitude
of inerton cloud of an atom:   $\Lambda \simeq 2.4 \times 10^{-5}$  m, that
essentially exceeds the lattice constant $g_0 \simeq 4 \times 10^{-10}$
m.  Thus, hypothetical inerton clouds of atoms noticeably overlap in a
solid and it  is obvious that this overlap should  make  a  definite
contribution to the collective behaviour of atoms. Let us study this
problem in the context of the standard model of the harmonic
interaction of atoms in the crystal lattice.

        We will proceed from the Lagrangian
\begin{eqnarray}
L&=&\frac {M}{2}\sum_{\l\alpha}\dot\xi^2_{\l\alpha}  - \frac 12
{\sum_{\vec l \alpha,  \vec n \beta}}^{\prime}
V_{\alpha\beta}(\vec l -\vec n)\xi_{\vec l \alpha}\xi_{\vec n
\beta}          \nonumber      \\ &-&\sqrt{Mm}{\sum_{\vec l
\alpha,\vec n \beta}}^{\prime} \xi_{\vec l \alpha}
\tau^{-1}_{\alpha\beta}(\vec l-\vec n) \dot \chi_{\vec n \beta} +
\frac {m}{2}\sum_{\vec l \alpha} \dot \chi^2_{\vec l \alpha}.
\label{3}
\end{eqnarray}
Here the first two terms describe the vibrations of the sites in a
three-dimensional lattice (see, e.g. Davydov [23]) and the last
two terms, which we have additionally introduced, describe the
interaction  of the atoms with  inertons  and  the kinetic energy
of inertons.  $\xi_{\vec l \alpha}$  ($\alpha$ = 1, \ 2, \ 3) are
three components of atom displacement from the lattice site whose
equilibrium  position  is determined  by  the lattice vector $\vec
l$; \  $\dot \xi_{\vec l \alpha}$ are three components of the
velocity of this  atom; $V_{\alpha \beta} (\vec l - \vec n)$ are
the  components of  the elasticity  tensor of the   crystal
lattice.   Let $m$ be characteristic mass of the inerton cloud
and, if $\chi_{\vec l \alpha}$ ($\alpha$     = 1, 2, 3) are three
components of the  position of  the  inerton cloud for the atom
determined by the lattice vector $\vec l$,   then $\dot \chi_{\vec
l \alpha}$ are  three components of the velocity of this inerton
cloud.  $\tau^{-1}_{\alpha \beta} (\vec l - \vec n)$ are
components (generally they  might be tensor quantities) of the
rate of collisions between the inerton cloud of the atom
determined by the lattice vector $\vec n$ and  with the atom whose
equilibrium position is determined by  the  vector $\vec l$. The
prime at  the sum  symbol means  that  terms  with coinciding
indices $\vec l$  and $\vec n$ are not taken  into  account  in
summation.

  In a standard way, we carry out canonical transformations in (3)
with respect to collective variables,
both for atoms ($A_{\vec k} =  (A_{-\vec k})^{\ast}$)  and for
inerton clouds ($a_{\vec k} = (a_{-\vec k})^{\ast} $):
\begin{equation}
\xi_{\vec l \alpha}=\frac 1{\sqrt{NM}}\sum_{\vec k}e_{\alpha}A_{\vec k}
e^{i\vec k \vec l};
\label{4}
\end{equation}
\begin{equation}
\chi_{\vec l \alpha}=\frac 1{\sqrt{Nm}} \sum_{\vec k}e_{\alpha}a_{\vec k}
 e^{i\vec k \vec l}
\label{5}
\end{equation}
where  $e_{\alpha} \equiv e_{\alpha}(\vec k)$ are components of
the polarisation vector  and $N$ is the number of  atoms  in  the
crystal. On rearrangement, the Lagrangian (3) takes the form
\begin{eqnarray}
L= \frac 12
\sum_{\vec k \alpha}e_{\alpha}\dot A_{\vec k}e_{\alpha}\dot A _{-\vec k}
&-&\frac 12 \sum_{\vec k \alpha \beta} \tilde{V} _{\alpha \beta}
(\vec k)e_{\alpha} A_{\vec k}e_{\beta} A_{-\vec k}          \nonumber    \\
&-&\sum_{\vec k \alpha\beta} \tilde\tau^{-1}_{\alpha\beta}
 e_{\alpha}A_{\vec k}e_{\beta}\dot a_{-\vec k}
+ \frac 12 \sum_{\vec k \alpha}e_{\alpha} \dot a_{\vec k}e_{\alpha}
\dot a_{-\vec k}
\label{6}
\end{eqnarray}
where real elements of force matrices are
\begin{equation}
\tilde V_{\alpha\beta}(\vec k)=
\frac 1{M}\sum_{\vec l}V_{\alpha\beta}(\vec l)e^{i\vec k\vec l};
\label{7}
\end{equation}
\begin{equation}
\tilde{\tau}^{-1}_{\alpha\beta}(\vec k)=
\sum_{\vec l}\tau^{-1}_{\alpha\beta}(\vec l)e^{i\vec k \vec l}.
\label{8}
\end{equation}
Euler-Lagrange equations
$$
  \frac {d}{dt} \Bigl(\frac {\partial L}{\partial {\dot Q}_s} \Bigr)
                - \frac {\partial L}{\partial Q_s} = 0
$$
for the variables $Q_1=e_{\alpha}A_{\vec k}$ and $Q_2=e_{\alpha}a_{\vec k}$
are respectively
\begin{equation}
 e_{\alpha}\ddot A_{-\vec k} +
  \sum_{\beta}[\tilde{V}_{\alpha\beta}(\vec k)e_{\beta}A_{-\vec k} +
\tilde{\tau} ^{-1}_{\alpha\beta}(\vec k) e_{\beta} \dot a_{\vec k} ]  = 0;
\label{9}
\end{equation}
\begin{equation}
e_{\beta} \ddot a_{-\vec k} -
\sum_{\alpha}e_{\alpha}\tilde{\tau}^{-1}_{\alpha \beta}(\vec k)
\dot A_{\vec k} = 0.
\label{10}
\end{equation}
Differentiating Eq. (9) with respect to time and replace
(-$\vec k$)    for $\vec k$ we obtain
\begin{equation}
e_{\alpha}{\mathop{A}\limits^{...}}_{\vec k} +  \sum_{\beta}
[\tilde { V}_{\alpha \beta}(\vec k)e_{\beta}\dot  A_{\vec k}
+ \tilde { \tau}^{-1}_{\alpha \beta}(\vec k) e_{\beta}\ddot a_{-\vec
k}] = 0.  \label{11}
\end{equation}
Substituting  $e_{\beta} \ddot a_{-\vec k}$
from Eq. (10) into Eq.  (11), we  gain  the equation for
$A_{\vec k}$ which after integration over $t$  changes to
\begin{equation}
e_{\alpha} \ddot A_{\vec k}
+ \sum_{\beta}W_{\alpha \beta} (\vec k) e_{\beta} A_{\vec  k} = C
\label{12}
\end{equation}
here $C$ is the  integration constant and the force matrix
\begin{equation}
W_{\alpha\beta}(\vec k)
= \tilde V_{\alpha \beta}(\vec k)
+ \tilde \tau^{-1}_{\alpha \beta} \sum_{\alpha^\prime}
\tilde \tau^{-1}_{\alpha^\prime \beta} (\vec k)
\frac {e_{\alpha^\prime}}{e_{\beta}}.
\label{13}
\end{equation}
Eq. (12) has the form  of  a  standard  equation for collective
variables  of  the  crystal  lattice   and  it determines three
frequencies $\Omega_{s} (\vec k)$   ($s$ = 1, 2, 3), that is,
three  branches of acoustic vibrations: one longitudinal branch
(along $\vec k$) and two transverse ones  (normal to  $\vec k$).
The equation for the frequencies $\Omega_{s}(\vec k)$ has the form
\begin{equation}
|| \Omega^2_s (\vec k) - W_{\alpha \beta}(\vec k) ||
 = 0.
\label{14}
\end{equation}
But in our case, as seen from (13) the force matrix $W (\vec k )$
comprises in addition  to the elastic (electromagnetic nature)
component $\tilde V (\vec k)$ also the inerton component caused by
overlapping the inerton cloud  of  each atom with the adjacent
atoms and proportional to $(\tau^{-1})^2$.

     It seems likely that in a crystal under ordinary  conditions the
correction $(\tau^{-1})^2$ to the elastic matrix $V(\vec k)$ is small.
But given a sufficiently intensive external source of inertons,
this correction can substantially increase and then the inerton component
can  show  up explicitly.  Indeed,  given  an external inerton source,
Eq. (10) is replaced by the  generalised equation
\begin{equation}
e_{\beta}\ddot a_{-\vec k} - \sum_{\alpha}e_{\alpha}\tilde
{\tau}^{-1}_{\alpha\beta}(\vec k)\dot A_{\vec k}
= f_{\vec k \beta} \cos (\omega_{\vec k}t).
\label{15}
\end{equation}
With the permanently acting source, when the external force
$f_{\vec k}> \tilde {\tau}^{-1} \dot A_{\vec k}$, the equation
\begin{equation}
e_{\beta}\ddot a_{-\vec k} \simeq f_{\vec k\beta} \cos(\omega_{\vec k}t)
\label{16}
\end{equation}
follows from (15). Integrating (16)  over $t$ and then substituting
 $e_{\beta}\dot a_{-\vec k}$ from (16) into (9),  we obtain the equation
\begin{equation}
e_{\alpha}\ddot A_{\vec k} +
\sum_{\beta}\tilde{V}_{\alpha \beta}(\vec k)e_{\beta}A_{\vec k} =
\sum_{\beta}\frac {\tilde \tau_{\alpha\beta}^{-1}(\vec k)}
{\omega_{\vec k}}f_{\vec k} \sin (\omega_{\vec k}t).
\label{17}
\end{equation}
At a sufficiently large value of the  permanent disturbance, e.g. along
the projection $e_1$, it is easy to find from (17) the amplitude
$A^{(0)}_{k_{1}}$  of collective vibrations of atoms:
\begin{equation}
A^{(0)}_{k_{ 1}} = f_{k_{ 1}}
\frac {\tilde \tau^{-1}_{11}/ \omega_{\vec k}}
{\Omega^2 (\vec k) - \omega_{\vec k}^2}
\label{18}
\end{equation}
and, therefore, the amplitude of individual atom vibration
\begin{equation}
\xi _{n_{ 1}} = \frac 1{\sqrt{NM}} \sum_{k_{1}} A^{(0)}_{k_{ 1}}
e^{i k_{1}n_{1}}
\label{19}
\end{equation}
(as is generally known, the consideration of friction $\eta$
enables the limitedness and constant sign  of  amplitude
$A^{(0)}_{k_{1}}$ in (18)).  Thus, it is easily seen from (19) and
(18) that given  disturbing force $f_{\vec k}$ of the applied
inerton field, the amplitude of atom vibrations in  the crystal
should increase, especially at resonance.

\vspace{4mm}

\section{Source of inertons}
\vspace{2mm} \hspace*{\parindent}
     A condensed medium and, in particular, our planet itself
could be considered as a source of inerton.
Moreover, two types of stationary inerton flows can be set off in
the terrestrial globe; their availability being  associated  with
the motion of the Earth: 1) the  orbital  motion  around
the Sun with the velocity $v_{01} \simeq $ 30 km/s \  and
2)  the proper rotation; with this motion  the  velocity  changes from
$v_{02}=0$ in the centre of the Earth  to $v_{02}=2\pi R_{\rm Earth}
/ 24\ {\rm hour} \simeq  $ 462 m/s in the equator surface.

   Actually, the motion of atoms of the Earth considered to be an ideal
globe moving as a single unit apparently does not differ in
principle from the motion of a free particle [18-20]. Structural
bonds, which keep atoms in the globe, lead to the coherence of
their motion. If we assume that the mean mass of atoms of the
Earth is  $M = 30 M_p$, then the de Broglie wavelength for the two
types of the  motion  will  be $\lambda
 _1  = h /Mv_{01} \approx  4 \times 10^{-13}$  m  and $\lambda _2 =
h/Mv_{02} \approx  1.5 \times 10^{-11}$ m respectively.  Substituting
 the values $\lambda _{1(2)}$ and $v_{01(02)}$ into formula (1), we
acquire the amplitudes of inerton clouds of the moving atoms of
the terrestrial globe: $\Lambda_1 \approx  8 \times 10^{-9}$ m and
$\Lambda_2 \approx 4 \times 10^{-5}$ m.  In both cases the overlap
of inerton clouds is substantial ($\Lambda _1 /g_0  \sim 10$  and
$\Lambda_2 /g_0  \sim 10^4$), but due to inequality $\Lambda_2 \gg
\Lambda_1 $ the degree of coherence between atoms is greater along
the velocity vector  ${\vec v}_{02}$  (i.e., along the West-East
line) than along  the  orbital velocity vector ${\vec v}_{01}$.

     Deviations from coherence in the motion of atoms caused by thermal
fluctuations and various mechanical,  physical and chemical
processes produce excitation of the atoms and as a result
generation of acoustic waves takes place. As a consequence
corresponding excitation of inertons (inerton waves) accompanying
the acoustic waves will appear as well. We may expect that owing
to the inequality $\Lambda_2 \gg \Lambda_1$ the generated inerton
waves will have a maximum intensity along the West-East line. The
velocity of the generated inerton wave propagation may be equal to
(or even exceed) the speed of light.

\vspace{4mm}

\section{Resonator }

\vspace{2mm} \hspace*{\parindent}
      If inerton waves really exist, then we can  try  to  amplify
their intensity in a resonator and then  to  register  the  waves
experimentally. Let us consider characteristics which  a  resonator of
inerton waves of the Earth should possess. As mentioned above, we can
separate out two types  of  inerton  waves  propagating  in  the
terrestrial globe: \  1) radial waves propagation along the diameter
(in parallel and antiparallel with the orbital velocity vector
${\vec v}_{01}$ of the Earth) and \  2) tangential waves propagation
over the surface zone of the Earth along the equatorial West-East line
(i.e., along or against the vector of the  rotational  velocity
${\vec v}_{02}$ of the Earth on the equator).  In  the  former  case,
the inerton wave front travels a distance $L_{\rm rad} = 4 R_{\rm Earth}$
in  the cyclic period and in the second case $ L_{\rm tan} = 2 \pi
R_{\rm Earth}$. From these two expressions we obtain the relation
\begin{equation}
L_{\rm tan}/ L_{\rm rad} = \pi /2.
\label{20}
\end{equation}
Apparently, relation (20) also  characterises  the ratio between
the wavelengths of the tangential and radial $n$ths  harmonics.

    The time of passing of the mentioned distances by the front of
an inerton wave is equal to $L_{\rm tan}/c \approx 0.13$ s and
$L_{\rm rad}/c \approx 0.09$ s correspondingly where we take the
speed of light $c$ for the velocity of spreading of inerton waves.
So, if the lifetime of acoustic waves which are excited in the
directions $L_{\rm tan}$ and $L_{\rm rad}$ far exceeds these two
magnitude, for instance by ten times, then inerton waves
accompanying the acoustic waves can turn round and pass through
the Earth the same number of times to the moment when the acoustic
excitations are fully scattered or absorbed (and hence the
inertons waves scattered too).

    Let us assume that a material object is located in the globe
surface far from its poles.  The object has  linear dimensions
$l_{\rm tan }$ in the horizontal plane along  the  West-East  line
and $l_{\rm rad}$ in the vertical direction, that is radial one.
Now, if the above dimensions satisfy relation (20), that is
\begin{equation}
l_{\rm tan}/ l_{\rm rad} =  \pi /2,
\label{21}
\end{equation}
then this object can play the role  of  a  resonator  of  inerton
waves of the Earth since the object has a  form  similar  to  the
Earth sphere (in the limit $ l_{\rm tan}, l_{\rm rad} \ll  R_{\rm
Earth}$). Such  a  resonator could   amplify  inerton  waves,
which  have wavelength $l_{\rm tan}$ in the horizontal direction
and $l_{\rm rad}$ in the vertical one  and  could  also amplify
their harmonics.

\vspace{4mm}

\section {Experimental}
\vspace{2mm} \hspace*{\parindent}
        Of course, it is by no means a simple problem to  register the
amplification  of  the amplitude of atom  vibrations in  the specimen
in question  placed into the resonator, the  more so as we can say
{\it a priori} nothing about the figure  of merit  of  the resonator.
However, a conclusion about  the existence of inerton field can be made
from the resulting integral effect.

     The amplification of vibrations of the atoms of the test specimen
in the resonator under the effect of the  force  of the inerton
field can apparently  be considered  as  an  analogy  of  the
effect  of ultrasound or micro waves sound. This is evident from
expression (13): the total force matrix of a crystal is formed by
the two terms having the same rights. As is generally known (see,
e.g. Yavorskii and Detlaf [24]), the effect of destroying and
crushing  various structures and polishing surfaces is
characteristic of ultrasound.  Thus, in our case, the experiment
can be carried out to study the extent of changes in the
non-uniformity  surface of  the specimen (its polish, finish or
sharpening) which stayed in the resonator for some time.

    The resonator was made of two identical  rectangular plates of
organic glass (transparent in the visible optical spectral
band and with the dimensions of the plate  $20 \times 16.5\
{\rm cm}^2$ and the plate thickness  3 mm)  which  were  sharpened and
bonded together along one of the long sides. The angle of inclination
of each of the plates  with respect  to the horizontal  has made up $
52^\circ$ so that in the section perpendicular  to  the  line  of
bonding of the plates,  the resonator had the shape of a  triangle.
The dimensions of the base of the triangle $l_{\rm tan}$  = 20 cm  and of
the height $l_{\rm rad}$ = 12.7 cm   satisfy relation (21). The resonator
was placed on  a  polished wooden horizontal  surface. No objects were
found  at a radial distance of about 70-cm  from the  resonator.  The
upper edge of the resonator (the line of the plates bonding) was
oriented along the North-South line and, therefore, one plate  faced
the West and the other the East. A wooden support column
 4.5  cm height, with the  cross-section area
$1 \times 1 \  {\rm cm}^2$ was placed in the center of the resonator
and the specimen being tested was put on the column.

     Our  investigation  was  related  with  the cutting  edge (point) of
a razor blade, but prior to putting it into resonator, a small
reference specimen in area  $15 \times 7 \  {\rm mm}^2$ was cut
out of the blade (the first dimension is the length  of  the
cutting  edge). The blade was put on  the column in the resonator
so  that the axis  of the blade was oriented along the North-South
line. The  main  action on the blade on the part of inerton waves
was  expected  in  the plane of the blade along the West-East line
(this action amounts to a peculiar sharpening of the cutting edge)
and the less  intensive action --  along vertical direction.
Razor blades  produced  by  four different  companies have been
studied:   "Shick"   (the Netherlands), "BIC" (Hungary), "Sputnik"
(Russia) and "Gillette" (the U.K.). Investigation of the structure
of the cutting edge  point of the reference specimen and of the
specimen  subjected  to  the hypothetical  inerton field was
carried out by scanning electron microscope JSM-35 (Japan)
operated in secondary electron mode under  25 kV accelerated
voltage. As for the blades  of the first three companies the time
of exposure of test specimens in the resonator varied from one to
two weeks. Nevertheless no substantial  distinction  has been
observed. In this connection the  exposure  time  was  increased
up to  30  days.

       Fragments of the cutting edge of a "Gillette"
blade are presented  for comparison  in  Fig. 1 (the reference and
test specimens, micrographs $a$ and $b$, respectively).  The same
for one more  "Gillette" blade, Fig. 2 (the reference ($a$) and
test ($b$) specimens). The Figs. 1 and 2 show that the fine
structure well discernible on the reference specimen (Figs. 1$a$
and 2$a$) is substantially smothered on the edge of the blade
which has stayed in the resonator for a month (Figs. 1$b$,  2$b$).
The morphologically more coarse structure is well preserved. (Note
that pressure, temperature, humidity, etc. could not make any
changes in the morphological structure of the test metal specimen
separated from the reference one by 1.5 meters; the two specimens
were found under the same atmospheric conditions.)
\begin{figure}
\begin{center}
\includegraphics[scale=1]{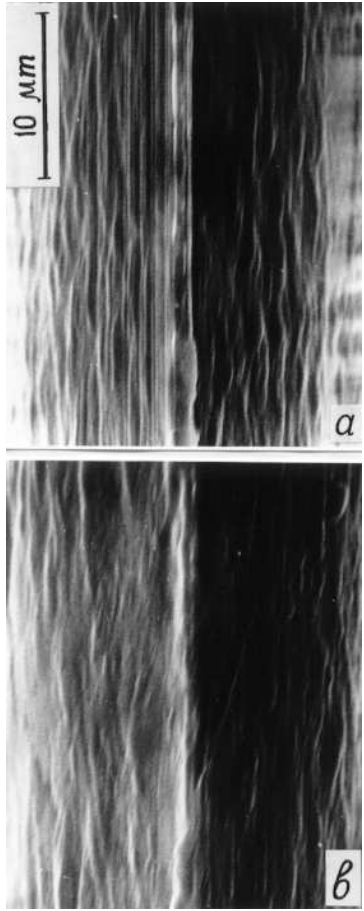}
\caption{ Micrographs of the cutting edge of the "Gillette" blade;
top view of the edge [3000$\times$]: \  $a$  - reference specimen;
\  $b$  - test specimen.} \label{Figure 1}
\end{center}
\end{figure}
\begin{figure}
\begin{center}
\includegraphics[scale=1]{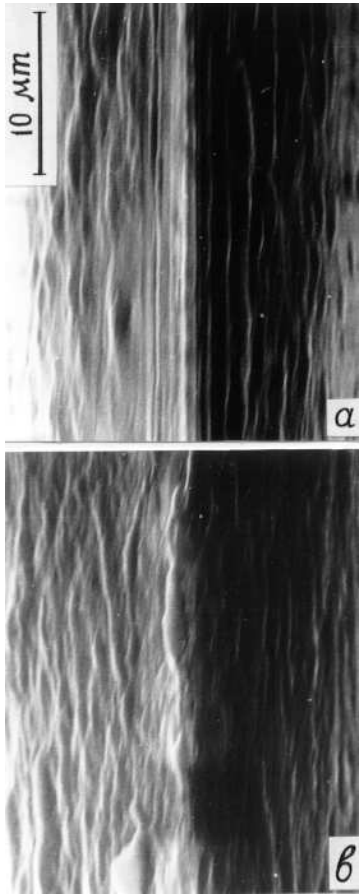}
\caption{ Micrographs of the cutting edge  of  one more "Gillette"
blade [3000$\times$]: \   $a$  - reference specimen; \  $b$  -
test specimen. } \label{Figure 2}
\end{center}
\end{figure}

   It can be seen that the established difference in the microstructure
of the cutting edge of the razor blade confirms the qualitative
treatment performed in Section 2 regarding the increase of the
amplitude of atom vibrations under the external inerton field. In
fact, it is central tenet of our concept that the influence of the
acoustic or inerton field on any heterogeneity-crystallite of the
specimen is not fundamentally different.  There was no ultrasound
in our experiment and therefore changes revealed in the fine
structure of the specimens could arise only from the impact of
Earth's inerton field. From the above reasoning it is clear that a
macroscopic mechanism of these changes is identical to
peculiarities of the absorption of sound in polycrystalline
bodies, Landau and Lifshitz [25].  Thus, if the wavelength of the
sound $\lambda _{\rm sound}$ is large as compared to the  size $d$
of individual crystallites, then each crystallite being in the
field of this sound wave is subjected to an uniformly distributed
pressure.  However, the deformation arising in this case is
nonhomogeneous due to anisotropy of the crystallites and boundary
conditions  on  their contact surfaces. As can be seen from the
Figs. 1 and 2, the typical size $d$ of a crystallite in the blade
is approximately $(0.1 - 1)$ $\mu$m.  The acoustic frequency is
limited by the value $\nu_{\rm Debye}\sim 10^{13}$ Hz and owing to
our main supposition that acoustic waves generate inerton waves,
we should ascribe this frequency also to the latter (see
expressions (14) and (13)). Let us take the speed of light $c$ for
the velocity of spreading of an inerton wave. Then we can make an
estimation of the lower length of the inerton waves excited in the
Earth which propagate through the specimen and destroy its fine
morphological structure: $c/\nu_{\rm Debye}\approx 30$ $\mu$m. The
upper value of inerton wavelength was obviously limited by the
resonator dimensions -- approximately 12 cm. So the wave length
and frequency spectra of inerton field influenced the specimen in
the resonator were restricted by the following uncertainties:
   $$ 30 \
\ \mu{\rm m} < \lambda_{\rm iner } < 12 \ \ {\rm cm};
   $$
  $$
2.5 \times 10^9 \ \ {\rm Hz} < \nu_{\rm iner} < 10^{13} \ \ {\rm
Hz}.
  $$

          If the whole system is swung through $ 90^\circ$, i.e. the
planes of the resonator and the cutting edge of the blade are oriented
to the North and South, then no distinctions are observed between the
fine structure of the test and the reference specimens
after the blade has stayed in the resonator for 30 days (Fig. 3).
Thus points to the fact that generation of the coherent inerton field  of
the  Earth along the North-South direction is absent.
\begin{figure}
\begin{center}
\includegraphics{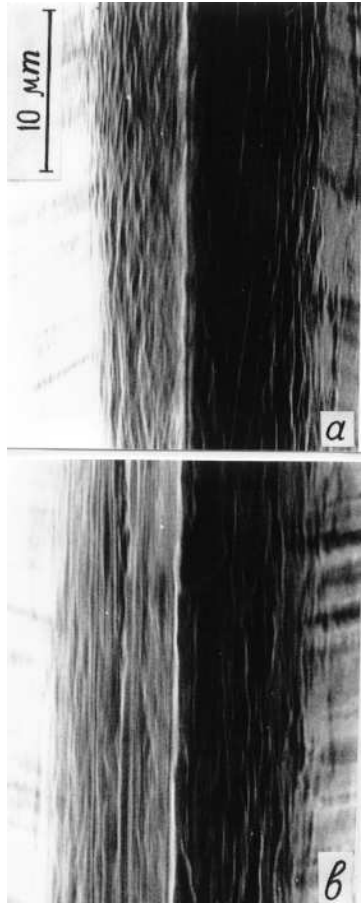}
\caption{Micrographs of the  reference  ($a$)
 and test ($b$) specimens ("Gillette"  blade) [3000$\times$]
 when the resonator planes and cutting edges of the blade
 are oriented to the North and South.} \label{Figure 3}
\end{center}
\end{figure}

\vspace{4mm}

\section{Concluding remarks}
\vspace{2mm} \hspace*{\parindent}
   The present research unambiguous demonstrates that a vacuum
should be considered in the form of a cellular elastic space. A
submicroscopic analysis of the behaviour of objects in space
performed in Refs. [18-20,22] and in this work has shown that any
motion of objects is accompanied by elementary excitations of
space called inertons.  It is obvious that these quasi-particles
should replace gravitons -- hypothetical massless particles of the
classical general theory of relativity. In fact, gravitons were
deduced on the assumption that a moving massive object did not
interact with the surrounding space (space was supposed to be
empty); besides, gravitons did not occupy any place in quantum
mechanics. By contrast, inertons fit naturally into quantum
mechanics explaining physical processes, which are veiled from
view by its formalism. Our data evidently display the existence of
inertons in the system studied.

    In such a manner inerton waves should be present in any system
consisting of a large number of bound particles. These waves are
excited and propagate in space and can influence material objects.
In particular, our planet itself is an inerton generator and,
evidently, the Earth inerton field may be considered as an
alternative to the aether wind which, speaking the language of
physicists of the nineteenth and early twentieth centuries, Born
[26], the Earth might experience in its motion through the world
aether.

Surprisingly, the mankind is familiar with the effect of influence
of the Earth inerton waves over a long time. Egyptian pyramids are
an glowing example. It is well-known (see, e.g. Schul and Pettit
[27]) that Egyptian pyramids and their small models possess
inexplicable properties:  the pyramids provide the mummification
of animal remains, depress germination of moistened grains, keep
up the razor blades sharp, {\it etc}.  The base of the pyramid is
a square oriented with a high degree of accuracy to the directions
of the world. The ratio between the side of the square $a_{\rm
pyr}$ and the height $h_{\rm pyr}$ of the pyramid satisfies
relation (21): $a_{\rm pyr}/h_{\rm pyr}= \pi /2$. This means that
like our resonator, which has the shape of a partly open book, the
Egyptian pyramid is a resonator for inerton waves generated by the
Earth as well.  By the way, the word {\it pyramid} means the
"inside fire" in Ancient Greek, that is the word itself alludes to
some inner properties of the pharaoh monument rather than points
to the monument habit.

Review by Puthoff and Targ [28] describes experiments on
transmission of mental information over a distance of hundreds
kilometers by extrasensitive participants placed into a special
metal room. The room shielded the participant-sender by two metal
screens; hence the room was absolutely impenetrable for the
electromagnetic field. Nevertheless the information transmitted
via the perceptive channel from one participant to another was
received successfully. However, it was pointed out in Ref. 28 that
the transmission was efficient only along the West-East line; but
it is just this direction that the intensity of inerton waves of
the Earth is maximum.  It is not ruled out that the inerton field
as an informational field plays an important role in other
phenomena of parapsychology (on parapsychology see, e.g. Dubrov
and Pushkin [29], and Morgan and Morris [30]).

     Of course the existence of inerton field should beyond any
reasonable doubt be sustained by other pure physical experiments.
Such candidates are in stock at present. Specifically, unusual
effects of the multiphoton ionisation of atoms of gas or
"effective photon", Panarella [31],  caused by low energy photons
provide a useful check on the availability of clouds of inertons
enclosing electrons [22]. Other remarkable unusual physical
effects revealed in the area of optical phenomena such as
diffraction/diffractionless of a single photon, Panarella [32],
photon as a particle-wave, Mizobuchi and Ohtake [33], and particle
tunneling and superluminal photonic tunneling, Nimtz and Heitmann
[34] can be easily understood from the submicroscopic view-point
drawing inertons which should be an integral part of the systems
studied and setups used by the researchers as well.

 \vspace{4mm}

{\bf Acknowledgement}

\vspace {2mm}

We thank greatly Professor Michel Bounias for the valuable remarks
and reprints of his articles cited herein.


\vspace {2mm}
\begin{thebibliography}{99}

\bibitem{1} Bergmann, P. G. -- Introduction to the theory of
            relativity, Gosudarstvennoe Izdatel'stvo Inostrannoy
            Literatury, Moscow, 1947, p. 251 (Russian translation).

\bibitem{2} Weinberg, S. -- {\it Gravitation and cosmology:
           principles applications of the general theory of
           relativity}, Mir, Moscow, 1975, p. 307 (Russian translation).

\bibitem{3} Einstein, A. -- \"Uber Gravitationwellen, {\it Sitzungsber.
            preuss. Akad. Wiss.} {\bf  1}, 154-167 (1918).

\bibitem{4} De Broglie, L. -- {\it Heisenberg's uncertainty relations
            and the probabilistic interpretation of wave mecanics}, Mir,
            Moscow, 1986 (Russian translation).

\bibitem{5} Stapp, H. P. -- Nonlocal character of quantum theory,
            {\it Am. J. Phys}. {\bf 65},  300-304 (1997).

\bibitem{6} Kar, K. C. -- {\it A new approach to the theory of
            relativity}, Bignar Kutir, Calcutta (1996).

\bibitem{7} Logunov, A. A. and  Metvirishvili, M. A. --
          {\it Relativistic theory of gravitation}, Nauka, Moscow (1989)
          (in Russian).

\bibitem{8} Hoyle, F. -- Mathematical theory of the origin of
            matter, {\it Astrophys. Space Sci.} {\bf 198}, 195-230 (1992).

\bibitem{9} Kohler, C. -- Point particles in 2+1 dimensional gravity
as defects in solid continua, Class. and Quant. Gravity, 12, pp.
L11-L15 (1995).

\bibitem{10} Bounias, M. -- {\it La cr\'eation de la vie: de la
            mati\`ere   \`a l'espirit, L'esprit et la mati\`ere}, \'Editions
            du Rocher, Jean-Paul Bertrand \`Editeur  (1990), p.
            38.

\bibitem{11} Bounias, M.  and Bonaly, A. -- On mathematical links
             between physical existence, observebility and information:
            towards a "theorem of something",  {\it Ultra Scientist of Phys.
            Sci.}, {\bf 6}, 251-259 (1994); \  Timeless space is provided
            by empty set, {\it ibid.} {\bf 8}, 66-71 (1996);
            -- On metric and scaling: physical co-ordinates in
            topological spaces, {\it Ind. J. Theor. Phys.} {\bf 44},
            303-321 (1996); \  Some theorems on the empty set as necessary
            and sufficient for the primary topological axioms of physical
            existence,  {\it Phys. Essays} {\bf 10}, 633-643 (1997).

\bibitem{12} Fomin, P. I. -- Zero cosmological constant and Planck
              scales phenomenology, in:  {\it Quantum Gravity. Proc. of IV
             Seminar on Quantum  Gravity, Moscow, USSR, May 1987}, eds.
             Markov, V.,  Berezin, V.  and  Frolov, V. P.,
             World Scientific Publishing Co., Singapore (1988) pp. 813-823.

\bibitem{13}  Aspden, H. -- The theory of gravitational constant,
               {\it Phys.  Essays} {\bf 2}, 173-179 (1989);
             -- The theory of antigravity, {\it ibid.} {\bf 4},
             13-19 (1996); --  {\it Aether Science Papers}, Subberton
           Publications, P.O. Box 35, Southampton SO16 7RB, England (1996).

\bibitem{14}  Haisch, B.,  Rueda, A.  and  Puthoff, H. E. --
            Inertia as a zero-point-field Lorentz force, {\it Phys. Rev. A}
            {\bf 49},  678-694 (1994).

\bibitem{15} Vegt, J. W. -- A particle-free model of matter based on
           electromagnetic self-confinement (III),  {\it Ann.  Fond. L. de
           Broglie} {\bf 21}, 481-506  (1996).


\bibitem{16} Winterberg, F. -- Physical continuum and the problem
            of a finistic quantum field theory, {\it Int. J. Theor. Phys}.,
            {\bf 32}, 261-277 (1993); -- Hierarchical order of Galilei and
            Lorentz invariance in the structure of matter, {\it ibid.},
            {\bf 32}, 1549-1561 (1993); -- Equivalence and gauge in
            the Planck-scale aether model, {\it ibid.} {\bf 34},
            265-285 (1995); -- Planck-mass-rotons  cold matter
            hyposesis, {\it ibid.} {\bf 34}, 399-409 (1995);
            -- Derivation of quantum mechanics from the Boltzmann equation
            for the Planck aether, {\it ibid.} {\bf 34}, 2145-2164
            (1995); -- Statistical mechanical interpretation of hole
            entropy, {\it Z.  Naturforsch.} {\bf 49a}, 1023-1030 (1994);
            Quantum mechanics derived from Boltzmann's equation
            for the Planck aether, {\it ibid.} {\bf 50a}, 601-605
            (1995); -- Planck scale physics and Newton's ultimate
            object conjecture, {\it ibid.} {\bf 52a}, 183-209 (1997).

\bibitem{17} Rothwarf, A. -- An aether model of the universe, {\it Phys.
             Essays} {\bf 11}, 444-466 (1998).

\bibitem{18} Krasnoholovets, V. and Ivanovsky, D. --  Motion of a
            particle and the vacuum, {\it Phys.  Essays} {\bf  6},
             554-563 (1993) (also http://xxx.lanl.gov  quant-ph 9910023).

\bibitem{18} Krasnoholovets, V. -- Motion of a relativistic
           particle and the vacuum, {\it Phys. Essays} {\bf 10}, 407-416
           (1997) (also http://xxx.lanl.gov  quant-ph 9903077).

\bibitem{19} Krasnoholovets, V. -- On the nature of spin, inertia
               and gravity of a moving canonical particle,
              {\it Ind. J. Theor. Phys.} {\bf 48}, no. 2 (2000), in press.

\bibitem{20} See Ref. [4], p.42.

\bibitem{21} Krasnoholovets, V. -- On the theory of the anomalous
             photoelectric effect stemming from a subsructure of
             matter waves, http:// xxx.lanl.gov quant-ph/9906091.

\bibitem{22} Davydov, A. S. -- {\it The theory of solids},  Nauka,
              Moscow (1976), p. 43 (in Russian).

\bibitem{23}  Yavorskii, B. M. and  Detlaf, A. A. -- {\it Handbook
              of physics}, Nauka, Moscow (1979), p. 558 (in Russian).

\bibitem{24}  Landau, L. D. and  Lifshitz, E. M., -- {\it Theory of
               elasticity}, Nauka, Moscow (1987), p. 182 (in Russian).

\bibitem{25}  Born, M. -- {\it Einstein's theory of relativity}, Mir,
              Moscow (1972), p. 209 (Russian translation).

\bibitem{26} Schul, B. and Pettit, E. -- {\it The secret power of
             pyramids}, Coronet, London (1978);  -- {\it The psichic
             power of pyramids}, Fawcett Publications,
             Inc.-Greenwich, Connecticut, (1976).

\bibitem{27} Puthoff, H. E. and Targ, R. --  A perceptual channel
            for information transfer over kilometer distances: historical
              perspective and recent research,  {\it Proc.  IEEE}
              {\bf 64}, 329-354 (1976).

\bibitem{28} Dubrov, A. P. and Pushkin, V. M. -- {\it Parapsychology
              and modern history}, Soviet-American enterprise
            "Sovaminco", Moscow, (1990) (in Russion).

\bibitem{29} Morgan, K. and Morris, R. L. --  A review of
              apparently successful methods for the enchancement of
             anomalous phenomena,  {\it J. Soc. Psychical Research}
             {\bf 58}, 1-9 (1991).

\bibitem{30} Panarella, E. -- Theory of laser-induced gas
            ionisation, {\it Found.  Phys.} {\bf 4}, 227-259 (1974);
            -- Effective photon hypothesis vs. quantum potential theory:
             theoretical predictions and experimental verification,
             in: {\it Quantum uncertainties. Recent and future
             experiments and interpretations. NATO ASI.  Series} {\bf B
             162}, {\it Physics}, eds.:  Honig, W. M., Kraft, D. W. and
             Panarella, E., Plenum Press, New York (1986),  pp. 237-269.

\bibitem{31} Panarella, E. -- Nonlinear behavior of light at very
             low intensities: the "photon clump" model, in: {\it Quantum
            uncertainties. Recent and future experiments and
            interpretations. NATO ASI. Series} {\bf B 162}, {\it Physics},
            eds.: Honig, W. M.,  Kraft, D. W. and Panarella, E.,
            Plenum Press, New York (1986), pp. 105-167.

\bibitem{32}  Mizobuchi, Yu. and  Ohtake, Yo. --  The duality of
         the  photon, {\it Nature} {\bf 367}, No. 6464, back cover (1994).

\bibitem{33} Nimtz, G. and  Heitmann, W. Superluminal photonic
            tunneling and quantum electronics, {\it Prog.  Quant.  Electr.}
            {\bf 21}, 81-108 (1997).
\end{thebibliography}
\end{document}